\documentclass[twocolumn,aps,floatfix,showpacs,showkeys]{revtex4}
\usepackage{color,graphicx}
\usepackage{dcolumn}
\usepackage{bm}
\usepackage{epsfig}
\usepackage{supertabular}
\usepackage[bookmarksopen]{hyperref}
\def\be {\begin{equation}}
\def\ee {\end{equation}}
\def\bea {\begin{eqnarray}}
\def\eea {\end{eqnarray}}
\def\nn {\nonumber}

\begin{document}

\title{Photons from jet - plasma interaction in relativistic heavy ion 
collisions}
\bigskip
\bigskip
\author{ Lusaka Bhattacharya}
\email{lusaka.bhattacharya@saha.ac.in}
\author{ Pradip Roy}
\email{pradipk.roy@saha.ac.in}
\affiliation{Saha Institute of Nuclear Physics, Kolkata - 700064, India}

\begin{abstract}
\leftskip1.0cm
\rightskip1.0cm
We expose the role of collisional energy loss on high $p_T$ photon
data measured by PHENIX collaboration by calculating photon yield
in jet plasma interaction. The phase space distribution of the 
participating jet is dynamically evolved by solving Fokker-Planck 
equation. It is shown that the  data is reasonably 
well reproduced when contributions from all the relevant sources are 
taken into account. Predictions at higher beam energies relevant 
for LHC experiment have been made.
\end{abstract}

\maketitle

\section{Introduction}

Heavy ion collisions have received significant attention in recent years. 
Various possible probes have been studied in order to detect the 
signature of QGP. Study of direct photon and dilepton 
spectra emanating from hot and dense matter formed in 
ultra-relativistic heavy ion collisions is a field of considerable 
current interest. Electromagnetic probes have been proposed to be one 
of the most promising tools to characterize the initial state of 
the collisions~\cite{jpr}. Because of the very nature of their interactions 
with the constituents of the system they tend to leave the system almost 
unscattered. In fact, photons (dilepton as well) 
can be used to determine the initial temperature, or equivalently the 
equilibration time. These are related to the final multiplicity of 
the produced hadrons in relativistic heavy ion collisions (HIC). By comparing 
the initial temperature with the transition temperature from lattice 
QCD, one can infer whether Quark Gluon Plasma (QGP) is formed or not.

There are various sources of photons from relativistic heavy ion collisions: 
(i) Direct photons are those which are produced in reactions of the
type $a\,b\,\rightarrow\, c\, \gamma$. 
One can subdivide this broad  category of "direct photons" into
"prompt", "pre-equilibrium", "thermal" (from QGP as well as hadronic 
phase) and finally the "jet-photon" also called "jet conversion photon" 
(photons from jet-plasma interaction) depending on their origin. 
(ii) Decay photons are basically the decay product of long lived 
secondaries 
($\pi^0 \to \gamma \gamma$, $\eta \to \gamma \gamma$, $\rho^0
\to \pi \pi \gamma$, $\omega \to \pi \gamma$ etc).
The first calculation of thermal photons from quark matter (QM) has
been done in Ref.~\cite{kapusta} using hard thermal loop (HTL). It is shown
that at a fixed temperature the contribution from QM is similar to 
that produced from hot hadronic matter. Here, our main concern is 
to calculate the jet-photons and compare them with the thermal photons 
from QM that has been calculated in Ref.~\cite{kapusta}.

The jet conversion mechanism~\cite{dks} occurs when a high energy jet 
interacts with the medium constituents via annihilation and Compton 
processes. It might be noted that this phenomenon (for Compton process) 
has been illustrated quite some time ago~\cite{pkrnpa} in the context 
of estimating photons from equilibrating plasma, where, it is assumed that 
because of the larger cross-section, gluons equilibrate faster providing 
a heat bath to the incoming quark-jet. A comparison of the 
photons, calculated in the above scenario, (equivalent to photons from 
jet-plasma interaction) with the direct photons (thermal) shows that the 
former remains dominant for photons with $p_T$ upto $6$ 
GeV~\cite{dks,pkrnpa}. It is to be noted that while 
evaluating jet-photon it is assumed in Ref.~\cite{dks} that the 
largest contribution to photons corresponds to 
$p_{\gamma} \sim p_q (p_{\bar q})$. This implies that the 
annihilating quark (anti-quark) directly converts into a photon. 
Moreover, the quark jet might lose energy due to scattering with 
the constituents of the thermal bath before participating in Compton and
annihilation processes. We include this effect on the photon productions.

Collisional energy loss of heavy fermion has been calculated long 
ago in Ref.~\cite{Braaten_PRD1, Braaten_PRD2} using HTL approach. 
The same for the light quarks and gluons has been discussed in 
Ref.~\cite{thomaplb} and revisited recently in Ref.~\cite{abhee05}. 
However, its importance was demonstrated in the context of RHIC in 
Ref.~\cite{roy06}. The measurements of non-photonic single electron 
data~\cite{phenixdil} show larger suppression than expected. These 
electrons mainly come from heavy quark decay where the radiative energy 
loss is suppressed due to dead cone effect. This observation has led to 
re-thinking the importance of collisional energy loss both for heavy as 
well as light quark~\cite{adil}. In view of this fact, collisional energy 
loss has been re-investigated in great 
detail~\cite{adil,jhep04,prd75054031,prc77044904} in recent times.

It is argued in Ref.~\cite{dks} that measurement of photons from
such a novel process can provide direct information about the
quark momentum. This is because of the assumption made in Ref.~\cite{dks} 
that photons are predominantly emitted at $p_{\gamma} \sim p_q$. This
implies that the thermal distribution of the participating parton is
evaluated at the photon momentum.
In this work we relax this assumption and calculate the photon yield 
from jet-plasma interaction. We consider photon production in the 
$p_T$ range $4 \leq p_T \leq 14 $GeV. It is to be noted that to produce 
such photons the required energy of the participating jet does not 
exceed (or remains below) the critical energy $E_c$ 
($E_{\rm rad} = E_{\rm coll}~{\rm at}~E=E_c$)~\cite{abhee05}. In fact, 
in this energy regime collisional loss seems to play important 
role~\cite{peshier,mustafa,qmjpg}. It is shown in Ref~\cite{mustafa,qmjpg}
that the quenching factor for high $p_T$ hadrons can be accommodated within
the framework of collisional energy loss only. 

Given the present scenario of energy loss mechanism in the context of 
RHIC data we, in this work, investigate the role of collisional energy loss
as calculated in Refs.~\cite{thomaplb,abhee05}. 
However, in order to see the effects of energy loss on jet-photon one 
should also incorporate the radiative energy loss and this has to be done 
in the same formalism in a realistic scenario. This has recently been done in
Ref.~\cite{prl100072301}, where it has been shown that the neutral pion
$p_T$ spectra is sensitive to the inclusion of collisional and radiative
energy loss.

In the photon production rate (from jet-plasma interaction) one of the 
collision partners is assumed to be in equilibrium and the other (the jet) 
is executing random motion in the heat bath provided by quarks 
(anti-quarks) and gluons. Furthermore, the interaction of the jet is 
dominated by small angle scattering. In such scenario the evolution 
of the jet phase space distribution is governed by Fokker-Planck 
(FP) equation where the collision integral is approximated by 
appropriately defined drag and diffusion coefficients.

The plan of the paper is as follows. We give a brief 
description of photon production from QGP in section IIA. The evolution of 
jet quark and photon $p_T$ distributions are discussed in sections 
IIB and IIC respectively. We then briefly mention the necessary formulae for 
photon production in initial hard collisions in section IID. 
Section III is devoted to the discussions of results and 
finally, we summarise in section IV.

\section{Formalism}
\subsection{Thermal Photon Rate}

The lowest order processes for photon emission from QGP are the
Compton scattering ($q ({\bar q})\,g\,\rightarrow\,q ({\bar q})\,
\gamma$) and annihilation ($q\,{\bar q}\,\rightarrow\,g\,\gamma$)
process. The total cross-section diverges in the limit $t$ 
or $u \to 0$. These singularities have to be shielded by thermal 
effects in order to obtain infrared safe calculations. It has been argued 
in Ref.~\cite{kajruus} that the intermediate quark  acquires a thermal 
mass in the medium, whereas the hard thermal loop (HTL) approach of 
Ref.~\cite{Brapi} shows that very soft modes are suppressed in a 
medium providing a natural cut-off $k_c \sim gT$.

We assume that the singularities can be shielded by the
introduction of thermal masses for the participating partons.
The differential cross-sections for Compton and
annihilation processes are given by~\cite{wong},
\begin{eqnarray}
&&\frac{d\sigma(qg\to q\gamma)}{d\hat t}=\frac{1}{6}(\frac{e_q}{e})^2\frac{8\pi
  \alpha_s \alpha_e}{(\hat s -m^2)^2}
\Big[(\frac{m^2}{\hat s -  m^2} +\frac{m^2}{\hat u -m^2})^2\nonumber\\
&+&(\frac{m^2}{\hat s -m^2}+\frac{m^2}{\hat u  -m^2})  
- \frac{1}{4}(\frac{\hat s -m^2}{\hat u- m^2}
+\frac{\hat u - m^2}{\hat s-m^2})\Big]
\end{eqnarray}
and
\begin{eqnarray}
&&\frac{d\sigma(q \bar q\to g \gamma)}{d\hat t}=-\frac{4}{9}
\frac{8\pi\alpha_s\alpha_e}{\hat s(\hat s-4m^2)}
\Big[(\frac{m^2}{\hat t-m^2}+\frac{m^2}{\hat u-m^2})^2
\nonumber\\
&+&(\frac{m^2}{\hat t-m^2}+\frac{m^2}{\hat u-m^2})
-\frac{1}{4}(\frac{\hat t-m^2}{\hat u-m^2}+\frac{\hat u-m^2}{\hat t-m^2})\Big]
\end{eqnarray}
where $m$ is the in-medium thermal quark mass. 
$m^2=2{m_{th}}^2=4\pi \alpha_s T^2/3$, $\alpha_e$ and $\alpha_s$ are the 
electromagnetic fine-structure constant and the strong interaction 
coupling constant, respectively. The static photon rate in
$1+2\rightarrow 3+\gamma$ can be written as~\cite{jpr,kapusta}
\begin{eqnarray}
\frac {dN^{\gamma}}{d^4xd^2p_T dy}&=&\frac{\mathcal {N}_i}{(2\pi)^7 E_\gamma}
\int d{\hat s} d{\hat t}
|\mathcal{M}_i|^2 \times\int dE_1 dE_2 \nonumber\\
&&\frac{f_1(E_1) f_2(E_2)(1 \pm f_3(E_3))}{\sqrt{a{E_2}^2+2bE_2+c}}
\label{rate}
\end{eqnarray}
\rm where \nonumber
\begin{eqnarray}
a&=&-(\hat s+\hat t-{m_2}^2-{m_3}^2)^2 \nonumber\\
b&=&E_1(\hat s+\hat t-{m_2}^2-{m_3}^2)({m_2}^2-\hat t)\nonumber\\
&+&E[(\hat s+\hat t-{m_2}^2-{m_3}^2)\nonumber\\
&\times&(\hat s-{m_1}^2-{m_2}^2)-2{m_1}^2({m_2}^2-\hat t)]\nonumber\\
c&=&{E_1}^2({m_2}^2-\hat t)^2-2E_1E[2{m_2}^2(\hat s+\hat t-{m_2}^2-{m_3}^2)
\nonumber\\
&-&({m_2}^2-\hat t)(\hat s-{m_1}^2-{m_2}^2)]\nonumber\\
&-&E^2[(\hat s-{m_1}^2-{m_2}^2)^2-4{m_1}^2{m_2}^2]\nonumber\\
&-&(\hat s+\hat t-{m_2}^2-{m_3}^2)({m_2}^2-\hat t)\nonumber\\
&\times&(\hat s-{m_1}^2-{m_2}^2)+{m_2}^2(\hat s+\hat t-{m_2}^2-{m_3}^2)^2
\nonumber\\ 
&+&{m_1}^2({m_1}^2-\hat t)^2\nonumber\\
E_{1,min}&=&\frac{\hat s+\hat t-{m_2}^2-{m_3}^2}{4E}\nonumber\\
&+&\frac{E{m_1}^2}{\hat s+\hat t-{m_2}^2-{m_3}^2} \nonumber\\
E_{2,min}&=&\frac{E{m_2}^2}{{m_2}^2-\hat t}+\frac{{m_2}^2-\hat t}{4E}
\nonumber\\
E_{2,max}&=&-\frac{b}{a}+\frac{\sqrt{b^2-ac}}{a}\nonumber
\end{eqnarray}
$ f_1 (E_1)$ , $f_{2}(E_2)$ and $f_3(E_3)$ are the distribution
functions of 1st, 2nd and 3rd parton respectively. $\hat s, \hat u$ and
$\hat t$ are the usual Mandelstam variables. $\mathcal {M}_i$ represents 
the amplitude for Compton or annihilation process. $\mathcal {N}_i$ is the 
overall degeneracy factor. For Compton scattering $\mathcal {N}_i=320/3$ 
and for annihilation process $\mathcal {N}_i=20$ when summing over 
u and d quarks.

\subsection {Fokker - Planck Equation: Parton transverse 
momentum spectra}

As mentioned already in the introduction that the quark jet here is 
not in equilibrium. Therefore the corresponding distribution function 
that appears in Eq.~(\ref{rate}) is calculated by solving the FP equation. 
The FP equation, can be derived from Boltzmann kinetic equation (BKE)if 
one of the partner of the binary collisions is in thermal equilibrium 
and the collisions are dominated by the small angle scattering 
involving soft momentum 
exchange~\cite{roy06,alamprl94,svetitsky,moore05,ducati,rajuprc01,rapp}. 

To arrive at the relevant FP equation from BKE we assume that there 
is no external force and therefore,
\bea
\left ( \frac{\partial }{\partial t} +
{\bf v_p\cdot \nabla_r} \right )
f({\bf p,x},t) =C[f({\bf p,x},t)]
\label{eq:boltz}
\eea
Here, quarks have a phase space distribution which evolves in
time and the collision term is evaluated by considering ultra-relativistic
scattering of the quarks and gluons which eventually are expressed in terms
of transport coefficients. For a Bjorken expansion~\cite{bj} the probability 
distribution is independent of the transverse coordinates and invariant
under boosts in the $z$-direction and Eq.~(\ref{eq:boltz}) is considerably 
simplified.
Considering that the system expands in the longitudinal direction
Eq.~(\ref{eq:boltz}) takes the following form \cite{baym}:
\bea
\frac{\partial f({\bf p},z,t)}{\partial t} +
v_{pz}\frac{\partial f({\bf p},z,t)}{\partial z} =C[f({\bf p},z,t)].
\eea
Here, $v_{pz}=p_z/E_p$ (for light partons $E_p=|p|$). This equation can
be simplified further for the central rapidity region which is boost
invariant in rapidity, which implies
\bea
f({\bf p_T},p_z,z,t)=f({\bf p_T},p_z^\prime,\tau).
\eea
Here $p_z^\prime=\gamma(p_z- u_z p)$, the transformation velocity
$u_z=z/t$, $\gamma=(1-u_z^2)^{-1/2}=t/\tau$ and $\tau=\sqrt{t^2-z^2}$
denotes the proper time. Using the Lorentz transformation relation
$\partial\tau/\partial z|_{z=0}=0$, $\gamma_{z=0}=1$  and
$\partial p_z^\prime/\partial z|_{z=0}= -p/t$, one finds
\bea
v_{pz}\frac{\partial f}{\partial z}=-\frac{p_z}{t}
\frac{\partial f}{\partial p_z}
\eea
Therefore the Boltzmann equation takes the following form
\bea
\frac{\partial f({\bf p_T},p_z,t)}{\partial t} |_{p_zt} =
\left (\frac{\partial}{\partial t}
-\frac{p_z}{t}\frac{\partial}{\partial p_z}\right )f({\bf p_T},p_z,t)
\eea
\bea
\left (\frac{\partial}{\partial t}
-\frac{p_z}{t}\frac{\partial}{\partial p_z}\right )f({\bf p_T},p_z,t)
=C[f({\bf p_T},p_z,t)].
\label{eq:boltzBj}
\eea
Evidently in Eq.~(\ref{eq:boltzBj}), the second term on the left hand side
represents the expansion while the right hand side characterizes the 
collisions. The latter can be written in terms of the differential 
collision rate $W_{{\bf p,q}}$
\bea
C[f({\bf p_T},p_z,t)]=\int d^3q 
[W_{p+q; q} f({\bf p+q})-W_{p; q} f({\bf p})]
\eea
which quantifies the rate of change of the quark momentum from ${\bf p}$ to
${\bf p-q}$, $W_{{\bf p,q}}= d\Gamma({\bf p,q})/d^3q$, where
$\Gamma$ represents scattering rates.

In a partonic plasma, small angle collisions, with parametric dependence of
$O(g^2T)$, are more frequent than the large angle scattering rate. The
latter goes as $\sim O(g^4T)$.  Therefore the distribution function does not
change much over the mean time between two soft scatterings. This allows
us to approximate $f({\bf p+q}) \simeq f({\bf p})$. In contrast,
$W_{{\bf p,q}}$, being
sensitive to small momentum transfer, falls off very fast with increasing
$q$. Therefore, we write
\bea
W_{{\bf p+q,q}}f({\bf p + q})
\simeq
W_{{\bf p,q}}f({\bf p})
+ q_i \frac{\partial}{\partial p_i} (W_{\bf p,q}f)\nn\\
+\frac{1}{2} q_i q_j  \frac{\partial^2}{\partial p_i \partial p_j}
(W_{\bf p,q}f)
\eea
With these approximation, Eq.~\ref{eq:boltzBj} can be written as
\bea
\left (\frac{\partial}{\partial t}
-\frac{p_z}{t}\frac{\partial}{\partial p_z}\right )f({\bf p_T},p_z,t)
=\frac{\partial}{\partial p_i}A_i({\bf p}) f({\bf p}) +\nn\\
\frac{1}{2}
\frac{\partial}{\partial p_i \partial p_j}[B_{ij}({\bf p})f({\bf p})],
\label{fpexp}
\eea
where we have defined the following kernels,
\bea
A_i&=&\int d^3q W_{\bf p,q} q_i \\
B_{ij}&=&\int d^3q W_{\bf p,q} q_i q_j
\eea
Writing $A_i=p_i\eta$ and 
$B_{ij}=B_t (\delta_{ij}-\frac{p_i p_j}{p^2}) + B_l \frac{p_ip_j}{p^2}
$
we arrive at the following equation
\begin{eqnarray}
\left (\frac{\partial}{\partial t}
-\frac{p_z}{t}\frac{\partial}{\partial p_z}\right )f({\bf p},t)
=\frac{\partial}{\partial p_i}[p_i\eta f({\bf p},t)]\nonumber\\ 
+\frac{1}{2}
\frac{\partial^2}{\partial p_z^2 }[{B_l}({\bf p})
f({\bf p},t)]   
+\frac{1}{2}\frac{\partial^2}{\partial p_T^2}[{B_t}f({\bf p},t)]
\label{fpexp}
\end{eqnarray}
In Eq.~(\ref{fpexp}) $f({\bf p},t)$ represents the non-equilibrium 
distribution of the partons under study, $\eta=(1/E)dE/dx$, denotes 
drag coefficient,
$B_l=d\langle(\Delta p_z)^2\rangle/dt$,
$B_t=d\langle(\Delta p_T)^2\rangle/dt$, 
represent diffusion constants along parallel and perpendicular 
directions of the propagating partons.

The transport coefficients, $\eta$, $B_l$ and $B_t$
appeared in Eq.~(\ref{fpexp}) can be calculated from the
following expressions:
\begin{eqnarray}
\frac{dE}{dx}&=&\frac{\nu_i} {(2\pi)^5}
\int\frac{d^3kd^3qd\omega}{2k2k^\prime 2p 2p^\prime}
\delta(\omega-{\bf v_p\cdot q})\delta(\omega-{\bf v_k\cdot q})\nonumber\\
&\times&\langle {\cal M} \rangle_{t\rightarrow 0}^2
f(|{\bf k}|)\left[1\pm f(|{\bf k}+{\bf q}|)\right]\omega
\end{eqnarray}
\begin{eqnarray}
B_{t,l}&=&\frac{\nu_i }{(2\pi)^5}
\int\frac{d^3kd^3qd\omega}{2k2k^\prime 2p 2p^\prime}
\delta(\omega-{\bf v_p\cdot q})\delta(\omega-{\bf v_k\cdot q})\nonumber\\
&\times&\langle {\cal M} \rangle_{t\rightarrow 0}^2 
f(|{\bf k}|)\left[1\pm f(|{\bf k}+{\bf q}|)\right]q_{t,l}^2 
\end{eqnarray}
in the small angle limit~\cite{abhee05,roy06}.
Here $f(|{\bf k}|,t)$ denotes the thermal distributions for the
quarks (Fermi-Dirac) or gluons (Bose-Einstein) and $\nu_i$ stands for the
statistical degeneracy factor for the $i^{th}$ parton species. 
The  matrix elements required to calculate the transport 
coefficients include diagrams involving exchange of massless 
gluons which render $dE/dx$ and $B_{l,t}$ infrared 
divergent. Such divergences can naturally be cured by using 
HTL~\cite{Brapi} corrected propagator 
for the gluons, i.e. the divergence is shielded by plasma effects.

In the coulomb gauge the gluon propagator for the transverse 
and longitudinal modes are denoted by  $D_{00}=\Delta_{l}$ and
$D_{ij}=(\delta_{ij}-q^i q^j/q^2)\Delta_{t}$ with~\cite{lebellac}:
\begin{eqnarray}
\Delta_{l}(q_0,q)^{-1}&=&q^2-\frac{3}{2}\omega_p^2
\left [\frac{q_0}{q}ln\frac{q_0+q}{q_0-q}-2\right ]
\end{eqnarray}
\begin{eqnarray}
\Delta_{t}(q_0,q)^{-1}&=&q_0^2-q^2+\frac{3}{2}\omega_p^2\nonumber\\
&\times& \left [\frac{q_0(q_0^2-q^2)}{2q^3}
ln\frac{q_0+q}{q_0-q}-\frac{q_0^2}{q^2}\right ]
\end{eqnarray}
The HTL modified  matrix element in the limit of small angle scattering 
takes the following form \cite{abhee05,roy06} for all the partonic 
processes having dominant small angle contributions like $qg\rightarrow qg$,
$qq\rightarrow qq$ etc.:
\begin{eqnarray}
|{\cal{M}}|^2&=& g^4 C_{R} 16 (E_p E_k)^2\vert \Delta_l(q_0,q)  
\nonumber\\
&+& ({\bf{v_p}}\times {\hat{q}}).({\bf{v_k}} \times \hat{q})
\Delta_{t}(q_0,q)\vert^2
\end{eqnarray}
where $C_{R}$ is the appropriate color factor. With the screened interaction, 
the drag and diffusion constants can be calculated along the line of 
Ref.~\cite{roy06}. For jet with energy $E >> T$ Eqs.(5) and (6), in leading
log approximation, give (e.g. for $q\,q\rightarrow\,q\,q$)~\cite{abhee05} 
\begin{eqnarray}
\frac{dE}{dx}& = &\frac{\nu_q \pi \alpha_s^2 T^2}{6} 
\ln\left(\frac{E}{g^2T}\right)
\nonumber\\
B_{t}&= &\frac{2\nu_q \pi \alpha_s^2 T^3}{3} \ln\left(\frac{E}{g^2T}\right)
\nonumber\\
B_{l}&= &\frac{\nu_q \pi \alpha_s^2 T^3}{3} \ln\left(\frac{E}{g^2T}\right)
\end{eqnarray}
Similarly, drag and diffusion coefficients for the relevant processes 
can be calculated analogously. Having known the drag and diffusion, 
we solve the FP equation using Green's function techniques: 
If $P(\vec{p},t|\vec{p_0},t_i)$ is a solution to Eq.(\ref{fpexp}) 
with the initial condition
\begin{equation}
P(\vec{p},t=t_i|\vec{p_0},t_i) = \delta^{(3)}(\vec{p}-\vec{p_0})
\end{equation}
the full solution with an arbitrary initial condition can be 
obtained as~\cite{moore05} 
\begin{equation}
f(t,\vec p) = \int d^3{\vec p_0} P(\vec{p},t|\vec{p_0},t_i)f_0(\vec p_0)
\end{equation}
where for the initial condition $f(t=0, \vec p)=f_0(p_0)$ and
$P(\vec{p},t|\vec{p_0},t_i)$ is the Green's function of the partial
differential Eq.(15).

Thus, to obtain $p_T$ distribution of the jets at time
$t$ we need to convolve the Green's function with the
initial spectrum~\cite{rapp}:
\begin{equation}
E\frac{dN}{d^3p}(p_T,t) = \int\,d^3p_{0}\,
P(\vec{p},t|\vec{p_0},t_i) E_0\frac{dN}{d^3p_0}\,
\label{rapp_prc71}
\end{equation}
We use the initial parton $p_T$ distributions 
(at the formation time $t_i$) taken from~\cite{dks,muller}:
\begin{eqnarray}
\frac{dN}{d^2p_{0T}dy_0}|_{y_0=0}=
\frac{K N_0}{(1+\frac{p_{0T}}{\beta})^\alpha},
\end{eqnarray}
where $K$ is a phenomenological factor ($\sim 1.5 - 2$) which takes 
into account the higher order effects. The values of the parameters are 
listed in Table.~\ref{parameter}.
\begin{table}[h]
\begin{center}
\begin{tabular}{|c|c|c|c|c|}
\hline\hline
& \multicolumn{2}{|c|}{RHIC} & \multicolumn{2}{|c|}{LHC} \\\cline{2-5}
& $q$ & $\bar q$ & $q$ & $\bar q$ \\\hline\hline
 $N_0~[1/GeV^2]$ & $5.0\times 10^2$& $1.3\times 10^2$& $1.4\times 10^4$&
$1.4\times 10^5$ \\ \hline
$\beta~[GeV]$ & 1.6& 1.9& 0.61& 0.32 \\ \hline
$\alpha$ & 7.9& 8.9& 5.3& 5.2 \\\hline\hline
\end{tabular}
\end{center}
\caption{Parameters for initial parton $p_T$ distribution.} 
\label{parameter}
\end{table} 

\subsection{Space time evolution}

In this section we discuss how to obtain the space-time 
integrated rate of photons from jet plasma interaction 
using Bjorken hydrodynamics~\cite{bj}. Note that, for jet photon, 
one of the distribution function appearing in Eq.(3) should be 
replaced by the phase-space distribution of the incoming jet. 
We assume invariant Bjorken correlation~\cite{Lin} between the 
particle rapidity ($y$) and the space time rapidity ($\eta$) to 
obtain the phase phase distribution of the jet. Now we have 
\begin{eqnarray}
\nu_i\int \frac{d^3x d^3p}{(2 \pi)^3} f(x,p) = N_i
\end{eqnarray} 
where $N_i$ is the number of particles $i$ and $\nu_i$ is the spin-color
degeneracy. The phase-space distribution function for an incoming (quark) jet,
assuming Bjorken $\eta-y$ correlation~\cite{Lin}, is as follows,
\begin{eqnarray}
f_{jet}(\vec r,\vec p,t_i)  &=&
\frac{(2\pi)^3\mathcal{P}(r_{\perp})}{\nu_q \tau p_T}
\frac{dN}{d^2{p_T}dy}\delta(\eta-y) \nonumber\\
&=& \frac{(2\pi)^3\mathcal{P}(r_{\perp})t_i}{\nu_q \sqrt{{t_i}^2-{z_0}^2}}
\frac{p_T}{E^2}\frac{dN}{d^2{p_T}dy}\delta(z_0-v_z t_i)\nonumber\\
\label{jetp_1}
\end{eqnarray}
where, $t_i$ is the formation time of the jet, and $z_0$ is its 
position in the QGP expansion direction. We consider the jets to 
be massless. It is also assumed that the jets do not change 
direction due to its interaction with the plasma particles. In 
such case $f_{jet}$ can be factorized into a position space and 
a momentum space part and finally we obtain the phase space 
distribution of the jet at a later time $t^\prime$ and at $y=0$ 
as(see Ref.~\cite{prc72} for details),
\begin{eqnarray}
f_{jet}(\vec r,\vec p,t^{\prime})|_{y=0}&=&
\frac{(2\pi)^3\mathcal{P}(|{\vec w}_r|)~t_i}{\nu_q \sqrt{{t_i}^2-{z_0}^2}}
\frac{1}{p_{T}}\nonumber\\
&\times&\frac{dN}{d^2{p_{T}}dy}(p_T,t^{\prime})\delta(z_0)
\label{jetp}
\end{eqnarray}
where $\frac{dN}{d^2{p_{T}}dy}(p_T,t^{\prime})$ can be obtained
from Eq.(\ref{rapp_prc71}).
$\nu_q$ is the spin-color degeneracy factor for the 
incoming quark and $\mathcal{P}(|{\vec w}_r|)$ is the initial jet 
production probability distribution at the initial radial 
position ${\vec w}_r$ in the plane $z_0=0$, where
\begin{eqnarray}
|{\vec w}_r|&=&(\vec {r}-(t^{\prime}-t_i)~
\frac{\vec{p}}{\vec{|p|}})\cdot \hat{r}
\nonumber\\
&=&\sqrt{(r\cos{\phi}-t^{\prime})^2+r^2\sin^2{\phi}}  \ \ {\rm for}\ t_i\sim 0
\end{eqnarray}
and $\phi$ is the angle in the plane $z_0=0$ between the direction of the
photon and the position where this photon has been produced. 
We assume the plasma expands only longitudinally. Thus 
using $d^4x=rdrdt^{\prime}d\phi dz_0$ we obtain the transverse momentum
distribution (using Eqs.(\ref{rate}) and (\ref{jetp})) of photon as follows:
\begin{eqnarray}
\frac{dN^{\gamma}}{d^2p_Tdy}&=&\int d^4x ~ \frac{dN^{\gamma}}{d^4xd^2p_Tdy} 
\nonumber\\
&=&\frac{(2\pi)^3}{\nu_q}{\int_{t_i}}^{t_c}dt^{\prime}
{\int_0}^R rdr \int d\phi\mathcal{P}(\vec {w_r})\nonumber\\
&\times& \frac{{\mathcal{N}_i}}{16(2\pi)^7E_{\gamma}}\int 
d{\hat s}d{\hat t} |\mathcal{M}_i|^2\int {dE_1 dE_2}\nonumber\\
&\times& \frac{1}{p_{1T}}\frac{dN}{d^2 p_{1T}dy}(p_{1T},t^\prime)
\frac{f_{2}(E_2)
(1\pm f_3(E_3))}{\sqrt{a{E_2}^2+2bE_2+c}}\nn\\
\end{eqnarray}
where  $\frac{dN}{d^2 p_{1T}dy}(p_{1T},t^\prime)$ can be calculated
from Eq.(\ref{rapp_prc71}). $f_2,f_3$ are Fermi-Dirac or Bose-Einstein
distributions. $t_c$ is the time when phase transition from quark 
matter (QM) to hadronic matter (HM) begins and can be obtained by 
using Bjorken cooling law~\cite{bj}. 
$R$ is the transverse dimension of the system. $\phi$ dependence occurs 
only in $\mathcal{P}(\vec {w_r})$. So the $\phi$ integration can be done 
analytically as in Ref.~\cite{prc72}. The temperature profile is taken 
from Ref.~\cite{prc72}.
\vspace{0.6cm}
\begin{figure}[htb]
\centerline{\psfig{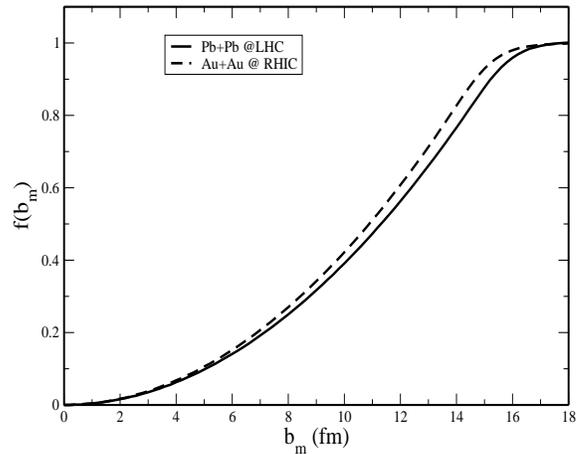}}
\caption{Centrality fraction as a function of maximum impact
parameter $b_m$.}
\label{fig1}
\end{figure}

\subsection{Hard Photons}

The large $p_T$-phenomenon in hadron-hadron collisions is well 
described by the perturbative QCD (pQCD) improved parton model. In this 
model it is assumed that the partonic structures of hadrons are revealed 
at high energies. The strong coupling constant $\alpha_s$ becomes `weak' 
so that perturbative expansion in powers of $\alpha_s$ becomes meaningful. 
Thus the partonic cross-section, reaction rates etc. can be calculated 
without much ambiguity and with great degree of accuracy. Therefore, within 
this model the hard photons coming from initial hard parton-parton collisions 
can be estimated very accurately. 

In order to calculate reaction of the type $h_A\,h_B\,\rightarrow\,\gamma\,X$
(where $h_A, h_B$ refer to hadrons), we assume that the energy is such that
the  partonic degrees of freedom become relevant and they behave incoherently.
The cross-section for this process can then be written in terms of
elementary parton-parton cross-section multiplied by the partonic flux
which depends on the parton distribution functions~\cite{ctq6pdf}.
The energy scale (so-called factorization scale) for this to happen is denoted
by $Q^2$, the square of the momentum transfer of the reaction. 
Starting with two body scattering at the partonic level the differential 
cross-section for the reaction of above type can be written as~\cite{hadpo}
\begin{eqnarray}
\frac{d\sigma_{\gamma,\rm hard}}{d^2p_Tdy} &=& 
K\,\sum_{abc} \int_{x_a^{\rm min}}^{1}\,dx_a\,
G_{a/h_A}(x_a,Q^2)\, G_{b/h_B}(x_b,Q^2)\nonumber\\
&\times&\,\frac{2}{\pi}\,\frac{x_a x_b}{2x_a-x_T {e^y}}
\frac{d{\sigma}}{d{\hat t}}(ab\rightarrow \gamma c).
\label{eq8}
\end{eqnarray}
where, $x_T=2p_T/\sqrt{s}$. The elementary partonic cross-sections for 
Compton scattering and annihilation process are given earlier.
\vspace{0.7cm}
\begin{figure}[htb]
\centerline{\psfig{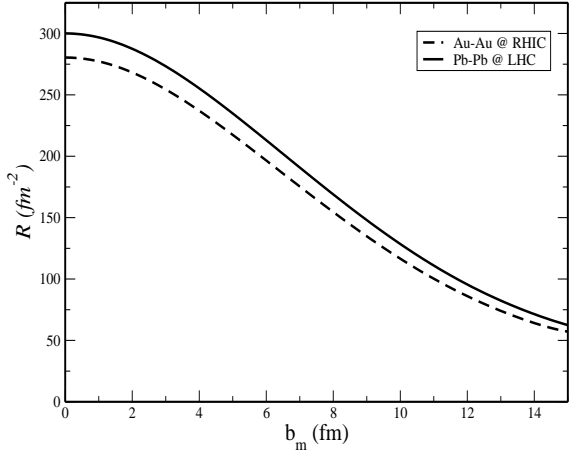}}
\caption{${\cal R}$ as a function of maximum impact
parameter $b_m$ corresponding to RHIC and LHC.}
\label{fig2}
\end{figure}
We also include photons from fragmentation process. This is accomplished by 
introducing the fragmentation function, $D_{\gamma /c}(z,Q^2)$, 
when multiplied by $dz$ gives the probability for obtaining a photon from 
parton $c$. Here $z$ is fractional momentum carried by the photon. The 
differential cross-section is, therefore~\cite{hadpo},
\begin{eqnarray}
\frac{d\sigma_{\gamma,\rm{frag}}}{d^2p_Tdy} & = & 
K\,\sum_{abcd}\, \int_{x_a^{\rm min}}^{1}\,dx_a\,\int_{x_b^{\rm min}}^{1}
G_{a/h_A}(x_a,Q^2)\,\nonumber\\ 
&\times&\,G_{b/h_B}(x_b,Q^2)\,D_{\gamma /c}(z,Q^2)\nonumber\\
&\times&\,\frac{1}{\pi z}\,
\frac{d{\sigma}}
{d{\hat t}}(ab\rightarrow cd),
\label{eq9}
\end{eqnarray}
where 
\begin{eqnarray}
z & = & \frac{x_T}{2x_b}e^{y}\nonumber\\
x_b^{\rm min} &=& \frac{x_a x_T e^{-y}}{2x_a-x_Te^y}\nonumber\\
x_a^{\rm min} &=& \frac{x_T e^{y}}{2-x_Te^{-y}}
\end{eqnarray}
\vspace{0.65cm}
\begin{figure}[htb]
\centerline{\psfig{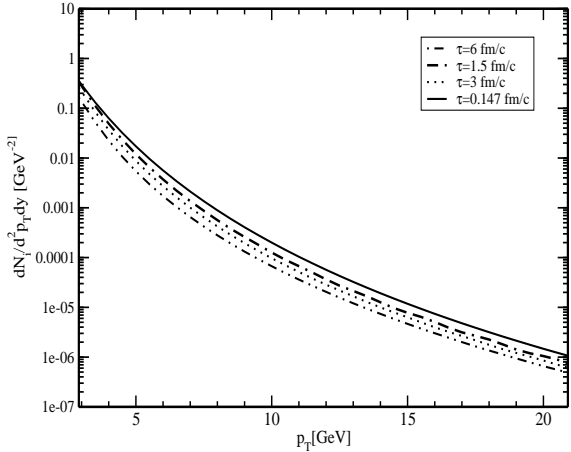}}
\caption{${p_T}$ distribution of light flavors. The parameters (such as 
initial temperature ${T_i}$ and time ${\tau_i}$) correspond to RHIC 
experiment. ${T_i}= 0.446$ GeV and ${\tau_i}= 0.147$ fm/c.}
\label{fig3}
\end{figure}
Once the photon production cross-section is obtained from hadron-hadron 
collision we can now determine the direct photon production rates due 
to hard scattering between partons from nucleus-nucleus collisions at 
relativistic energies. To do this we must note that the experimental data 
are given for a particular centrality. 
In order to take into account this experimental fact we 
introduce the centrality parameter (or the most inelastic fraction). 
It depends on the maximum impact parameter $b_m$ an can be calculated 
from the expression:
\begin{equation}
f(b_m) = \frac{\int_0^{b_m} d{\bf b}\,
\left(\frac{}{}1-[1-T_{\rm AB}(b)\sigma_{NN}^{in}]^{AB}\right)}
{\int_0^{\infty} d{\bf b}\,\left(\frac{}{}1-[1-T_{\rm AB}(b)\sigma_{NN}^{in}]^{AB}
\right)}
\label{eq17}
\end{equation}
From this expression we extract $b_m$ relevant for a given experiment and use 
it to calculate photons from initial hard collisions and from parton 
fragmentation. Thus the yield becomes
\begin{equation}
\frac{dN_{\rm AB}}{d^2p_T\,dy}(b_m) = {\cal R}(b_m)
\left[\frac{d\sigma_{\gamma,{\rm hard}}}{d^2p_T\,dy}+  
\frac{d\sigma_{\gamma,{\rm frag}}}{d^2p_T\,dy}\right]  
\label{eq18}
\end{equation}
where
\begin{equation} 
{\cal R}(b_m) \equiv \langle ABT_{\rm AB} \rangle=
 \frac{\int_0^{b_m}\,d^2{\bf b}\,AB\,T_{\rm AB}(b)}{\int_0^{b_m} d{\bf b}\,
\left(\frac{}{}1-[1-T_{\rm AB}(b)\sigma_{NN}^{in}]^{AB}
\right)} 
\label{eq19}
\end{equation}
and
\begin{equation}
T_{\rm AB}({\bf b}) = \int\,d^2{\bf s}\,T_{\rm A}({\bf s})\,T_{\rm B}(\bf b-s),
\label{eq13}
\end{equation}
is the nuclear overlap function.
\vspace{.2cm}
\begin{figure}[tb]
\centerline{\psfig{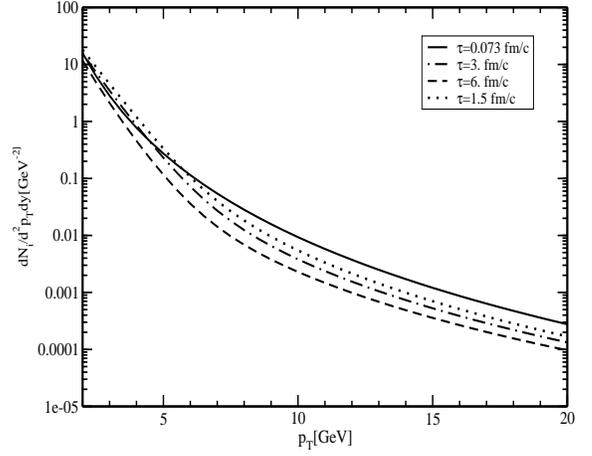}}
\caption{Same as Fig.(\protect\ref{fig3}) at LHC. ${T_i}= 0.897$GeV 
and ${\tau_i}= 0.073$ fm/c.}
\bigskip
\label{fig4}
\end{figure}

\section{Results}

First of all let us concentrate on the centrality measurements in an 
experiment. The photon measurement is done for a given centrality. For
example, $10\%$ centrality corresponds to $f(b_m)\sim 0.1$. 
In Fig.~\ref{fig1} we plot the most inelastic fraction as a function 
of the maximum impact parameter $b_m$ for Pb-Pb collisions at LHC. We 
obtain $b_m \sim 5$ fm for $10\%$ most central collision for Pb-Pb system. 
Similar value is obtained for RHIC. ${\mathcal R}$ vs $b_m$ given by 
Eq.~\ref{eq19} is plotted in Fig.~\ref{fig2} for Pb-Pb and Au-Au 
systems from which we obtain ${\cal R}\sim 215 (235) $ fm$^{-2}$ 
at RHIC (LHC) for $10\%$ centrality. We shall use these values 
while estimating hard photon yields at RHIC (LHC) energies.
%
\begin{figure}[tb]
\centerline{\epsfig{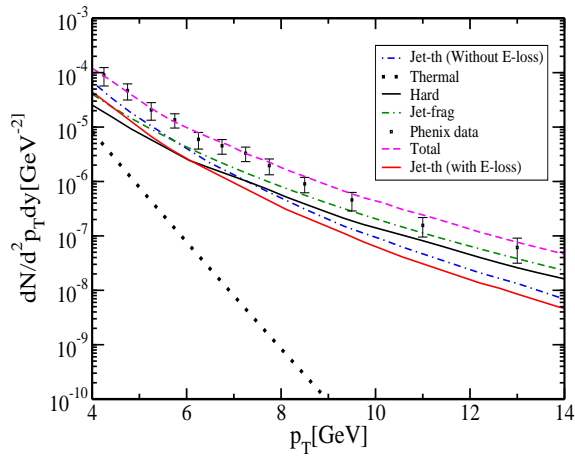}}
\caption{(Color online) $p_T$ distribution of photons at RHIC energies. 
The red (blue) curve denotes the photon yield from jet-plasma interaction 
with (without) energy loss.  The solid (dotted) black curve corresponds 
to hard (thermal) photons. The magenta represents the total yield 
compared with the Phenix measurements of photon data~\protect\cite{phenix}. 
${T_i}= 0.446$ GeV and ${t_i}= 0.147$ fm/c}
\label{fig5}
\end{figure}
We plot the transverse momentum distributions of quarks in Figs.~\ref{fig3}
and~\ref{fig4} for different times  at RHIC and LHC energies 
respectively where the initial distributions are taken from 
Eq.~(\ref{jetp}). It is seen that as the time increases the quark stays 
longer in the medium losing more energy. As a result the depletion in 
the distribution function is clearly revealed. It should be noted here 
that we do not include the induced gluon radiation which may further 
deplete the distribution at higher momenta.

In Fig.~\ref{fig5} we show the $p_T$ distribution of photons from various
processes which contribute at this high $p_T$ range. It is observed that due 
to the inclusion of energy loss in the jet-plasma interaction the yield is 
depleted. Our calculation without energy loss is similar to that in 
Ref.~\cite{dks} at RHIC energies. However, at LHC energies, as we shall see
below, the difference is by a factor of $2 - 3$. Thus, we see that the
assumption made in Ref~\cite{dks} could be valid for RHIC energies in the
$p_T$ range considered here, but at higher beam energies this is not a 
good approximation. 
It is observed from Fig.~\ref{fig5} due to the inclusion of energy 
loss the rate is lowered by a factor $\sim 1.5 (1.7)$ at $p_T= 4 (14)$. 
This is more or less similar to what is obtained in 
Ref.~{\cite{prc72}}. The total photon yield consists of jet-photon, 
photons from initial hard collisions, jet-fragmentation and 
thermal photons. It is seen that Phenix photon data is well 
reproduced in our model. At high $p_T$ region the data is 
marginally reproduced. The reason behind this is the 
following. For high $p_T$ photon the incoming jet must 
have high energy where the radiative loss starts to dominate. 
Inclusion of this mechanism might lead to a better description of the data 
in high $p_T$ range.
\vspace{0.7cm}
\begin{figure}[htb]
\centerline{\epsfig{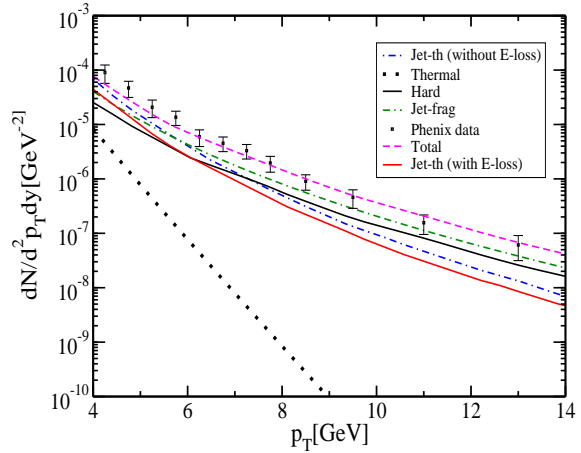}}
\caption{(Color online) Same as Fig.~(\protect\ref{fig5}) where in 
the total yield the contribution of photons from jet plasma interaction 
has been excluded.}
\bigskip
\label{fig6}
\end{figure}
In order to understand the role of photons from jet-plasma interaction
in describing the Phenix high $p_T$ photon data we show in Fig.~\ref{fig6}
where the contribution from jet-plasma interaction is excluded. It is
seen that in order reproduce the data with $3 < p_T < 6$ GeV one must
consider this extra source of photons.

\vspace{0.2cm}
\begin{figure}[htb]
\bigskip
\centerline{\epsfig{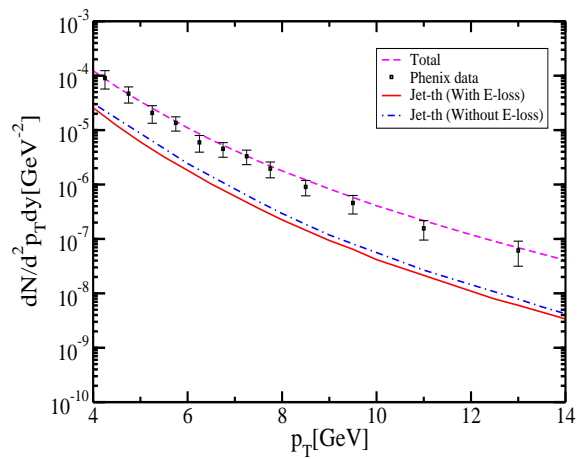}}
\caption{(Color online) $p_T$ distribution of photons at RHIC energies with 
$T_i= 0.350$ GeV and $\tau_i= 0.25$ fm/c. The magenta line corresponds
to total yield from all the sources as in Fig.~(\protect\ref{fig5}) with
energy loss included in the jet-photon contribution. }
\label{fig7}
\end{figure}

To cover the uncertainties in the initial conditions for a given beam 
energy, we consider another set of initial conditions at a lower 
temperature $T_i=0.350$ GeV and somewhat later initial time of 
$\tau_i=0.25$ fm/c. The yield for this set is shown in Fig.~\ref{fig7}. 
We see that the data is reproduced reasonably well. The inclusion of 
radiative energy loss will improve the situation further.

As mentioned before, we also consider the high $p_T$ photon production 
at LHC energies. The contributions from various sources are shown 
in Fig.~\ref{fig7}. Since the initial temperature in this case is 
higher, the plasma lives for longer time. Thus the energy loss suffered 
by the parton is more. As a result, the difference between the cases with 
and without energy loss is slightly more than what is obtained at RHIC.
\vspace{0.1cm}
\begin{figure}[htb]
\bigskip
\centerline{\epsfig{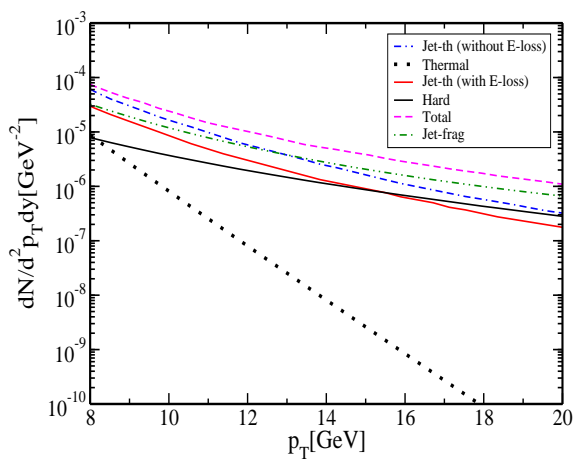}}
\caption{(Colour online) Same as Fig.(\protect\ref{fig5}) at LHC energies. 
$T_i= 0.897$ GeV and $\tau_i= 0.073$ fm/c.}
\label{fig8}
\end{figure}
We do not consider strange quark as it is Boltzmann suppressed because
of its high mass. However, if one considers strange quark the
effective number of flavours is taken as $2.5$. 
For $N_F = 2 (2.5)$ the lifetime of the plasma $(t_c - t_i)$ is of the 
order of $1.9 (2.6)$ $fm/c$ (the values of $T_c$ are taken from 
Ref.~\cite{braun}). So we see that the life of the plasma increases
marginally. As a result minor enhancement in the yield is expected.
On the other hand, the degeneracy factor ${\cal N}_i$ is $20 (30)$ 
$N_F = 2 (2.5)$ in case of annihilation proceses. So the
yield will be larger by a factor of the order of $1.5$. 
Similar thing happens in the case of compton scattering.

\section{Summary}

We have calculated the transverse momentum distribution of
photons from jet plasma interaction with collisional energy loss. 
It is shown that the assumption made in Ref.~\cite{dks} while calculating
photons from jet - plasma interactions may not be good at LHC energies (the
difference is by a factor of $2 - 3$). 
Phenix photon data have been contrasted with the present 
calculation and the data seem to have been reproduced well. We note that the
yield of thermal photons as well as photons from jet-plasma interaction is
very sensitive to initial temperature $(T_i)$ time $(\tau_i)$ and the equation
of state. It is shown that in order to describe the Phenix data in the domain 
$4 < p_T < 14$ GeV the contribution from jet-plasma interaction is 
found to be important. The energy of the jet quark to produce photons 
in this range is such that collisional energy loss plays a dominant 
role here. In view of this fact we have considered collisional energy loss
only. However, one should include both mechanisms in a 
consistent manner to describe the high $p_T$ data beyond $14$ GeV or so. 
We note that the data is over-predicted unless the energy loss is 
included. As we validate our model through the description of Phenix 
data we also predict the high $p_T$ photon yield that might be expected 
in the future experiment at LHC. We observe that the contribution 
from jet-plasma interaction is slightly more reduced as compared to 
the RHIC case. We notice that the inclusion of the radiative energy 
loss will further reduce the yield at high $p_T$. That is expected to 
give a better description of the Phenix data in the high energy regime.
 
Finally, it should be noted that the fragmentation photon should, 
in principle, be calculated taking into account the energy loss of 
the fragmenting parton while traversing the plasma. This can be 
done by calculating the modified $p_T$ distribution of the parton by 
using FP equation and then fragmenting it into photon as is done in 
case of high $p_T$ hadron production in relativistic heavy ion 
collisions~\cite{qmjpg}. The yield will be depleted as compared 
to the result shown in the present work. This can lead to the situation 
where jet-photons and fragmentation photons may be comparable in 
the intermediate $p_T$ domain as is evident from the present results. 
However, in the present work we do not consider 
this aspect as our main concern is to see the effect of collisional 
energy loss on photons from jet plasma interaction.


\noindent
\begin{thebibliography}{50}
\medskip


\bibitem{jpr}  J. Alam, S. Sarkar, P. Roy, T. Hatsuda and B. Sinha,
Ann. Phys. {\bf 286}, 159 (2000).

\bibitem{kapusta} J. I. Kapusta, P. Litchard and D. Seibert, 
Phys. Rev. D {\bf 44}, 2774 (1991). 

\bibitem{dks} R. J. Fries, B. Muller, and D. K. Srivastava, Phys. Rev. Lett. 
{\bf 90}, 132301 (2003).

\bibitem{pkrnpa} P. Roy, J. Alam, S. Sarkar, B. Sinha, and S. Raha, 
Nucl. Phys. A {\bf 624}, 687 (1997).


\bibitem{Braaten_PRD1} E. Braaten and M. H. Thoma, 
Phys. Rev. D {\bf 44}, 1298 (1991). 

\bibitem{Braaten_PRD2} E. Braaten and M. H. Thoma, 
Phys. Rev. D {\bf 44}, 2625 (1991). 

\bibitem{thomaplb} M. H. Thoma, Phys. Lett. B{\bf 273}, 128 (1991).

\bibitem{abhee05} A. K. Dutt-Mazumder, J. Alam, P. Roy, B. Sinha,
Phys. Rev. D {\bf 71}, 094016 (2005).

\bibitem{roy06} P. Roy, A. K. Dutt-Mazumder and J. Alam, Phys. Rev. C 
{\bf 73}, 044911 (2006).

\bibitem{phenixdil} S. S. Adler et al., Phenix Collaboration, Phys. Rev. 
Lett. {\bf 96}, 032301 (2006).

\bibitem{adil} A. Adil, M. Gyulassy, W. Horowitz and 
S. Wicks, Phy. Rev. C {\bf 75} 044906 (2007); M. Djordjevic, Phys. Rev. C {\bf
74} 064907 (2006); T. Renk, Phys. Rev. C {\bf 76} 064905 (2007).


\bibitem{jhep04} S. Peigne, P. B. Gossiaux, and T. Gousset, J. High 
energy Phys. {\bf 04}, 011 (2006).

\bibitem{prd75054031} J. Braun and H-J. Pirner, Phys. Rev. D {\bf 75}, 
054031 (2007).

\bibitem{prc77044904} A. Ayala, J. Magnin, L. M. Montano, and E. Rojas,
Phys. Rev. C{\bf 77}, 044904 (2008).

\bibitem{mustafa} M. G. Mustafa and M. H. Thoma, Acta Phys. Hung. A {\bf 22}, 
93 (2005). 

\bibitem{peshier} A. Peshier, Phys. Rev. C {\bf 75}, 034906 (2007).

\bibitem{qmjpg} P. Roy, J. Alam and A. K. Dutt-Mazumder, J. Phys. G {\bf 35}, 
104047 (2008).

\bibitem{prl100072301} G. Y. Qin, J. Ruppert, C. Gale, S. Jeon, G. Moore, and
M. G. Mustafa, Phys. Rev. Lett {\bf 100}, 072301 (2008).


\bibitem{kajruus} K. Kajantie and P. V. Russkanen Phys. Lett. {\bf B121}, 352
(1983).

\bibitem{Brapi} R. D. Pisarski and E. Braaten, Nucl. Phys. {\bf B337}, 569
(1990); {\it{ibid}} Nucl. Phys. {\bf B339}, 310 (1990).

\bibitem{wong} C. Y. Wong, Introduction to High Energy Heavy Ion Collisions,
1994, World Scientific, Singapore.

\bibitem{alamprl94} J. Alam, S. Raha and B. Sinha, Phys. Rev. Lett 
{\bf 73}, 1895 (1994).

\bibitem{svetitsky} B. Svetitsky, Phys. Rev. D {\bf 37}, 2484 (1988).


\bibitem{moore05} G. D. Moore and D. Teaney, Phys. Rev. C {\bf 71}, 
064904 (2005).

\bibitem{ducati} M. B. G. Ducati, V. P. Goncalves and L. F. Mackedanz,
hep-ph/0506241.


\bibitem{rajuprc01} J. Bjoraker and R. Venugopalan, Phys. Rev. C {\bf 63}, 
024609 (2001).

\bibitem{rapp} H. V. Hees and R. Rapp, Phys. Rev. C {\bf 71} 034907 (2005).

\bibitem{bj} J. D. Bjorken, Phys. Rev. D {\bf 27} 140 (1983).

\bibitem{baym} G. Baym, Phys. Lett. B {\bf 138} 18 (1984).

\bibitem{lebellac} M. Le Bellac, Thermal Field Theory, Cambridge 
University Press, Cambridge, 1996.

\bibitem{muller} B. Muller, Phys. Rev. C {\bf 67} 061901R (2003).

\bibitem{Lin} Z. Lin and M. Gyulassy, Phys. Rev. C {\bf 51}, 2177 (1995).


\bibitem{prc72} S. Turbide, C. Gale, S. Jeon and G. D. Moore, Phys. Rev. C 
{\bf 72} 014906 (2005).

\bibitem{ctq6pdf} J. Pumplin, D. R. Stump, J.Huston, H. L. Lai, P. Nadolsky, 
W. K. Tung, J. High Energy Phys. {\bf 012} 0207 (2002).

\bibitem{hadpo} J. F. Owens, Rev. Mod. Phys. {\bf 59} 465 (1987).

\bibitem{phenix}  S. S. Adler et al., Phys. Rev. Lett. {\bf 98} 012002 (2007) .

\bibitem{braun} J. Braun and H. Pirner, Phys. Rev. D {\bf 75} 054031 (2007).




\end{thebibliography}
\end{document}